\documentclass[english,aps,twocolumn,superscriptaddress,groupedaddress,nofootinbib]{revtex4}

\usepackage[T1]{fontenc}
\usepackage[latin1]{inputenc}
\usepackage{graphicx}
\usepackage{amsmath}
\usepackage{amssymb}
\usepackage{amsthm}
\usepackage{verbatim}
\usepackage{epic,eepic}
\usepackage{babel}
\usepackage{color}
\usepackage{bm}
\usepackage{mathtools}

\definecolor{mygreen}{RGB}{0, 122, 30}

\begin{document}

\widetext

\title{$0^+$ to $2^+$ neutrinoless double-$\beta$ decay of $^{76}$Ge, $^{82}$Se, $^{130}$Te and $^{136}$Xe in the microscopic interacting boson model}

\author{J. Ferretti}
\email{jacopo.ferretti80@gmail.com}
\affiliation{Department of Physics, University of Jyv\"askyl\"a, PO Box 35, FI-40014, Jyv\"askyl\"a, Finland}
\author{R. Maga\~na Vsevolodovna}
\email{Ruslan.Magana@ge.infn.it}
\affiliation{INFN, Sezione di Genova, via Dodecaneso 33, 16146 Genova (Italy)}
\author{J. Kotila}
\email{jenni.kotila@jyu.fi}
\affiliation{Department of Physics, University of Jyv\"askyl\"a, PO Box 35, FI-40014, Jyv\"askyl\"a, Finland}
\affiliation{Finnish Institute for Educational Research, University of Jyv\"askyl\"a, P.O. Box 35, FI-40014 Jyv\"askyl\"a, Finland}
\address{Center for Theoretical Physics, Sloane Physics Laboratory, Yale University, New Haven, Connecticut 06520-8120, USA}
\author{E. Santopinto}
\email{elena.santopinto@ge.infn.it}
\affiliation{INFN, Sezione di Genova, via Dodecaneso 33, 16146 Genova (Italy)}

\date{\today}

\begin{abstract}
Here, we study the neutrinoless double-$\beta$ ($0\nu\beta\beta$) decay between the ground state and the first $2^+$ state of $^{76}\mbox{Ge} \rightarrow {}^{76}\mbox{Se}$,  $^{82}\mbox{Se}  \rightarrow{}^{82}\mbox{Kr}$, $^{130}\mbox{Te} \rightarrow {}^{130}\mbox{Xe}$ and $^{136}\mbox{Xe} \rightarrow {}^{136}\mbox{Ba}$ systems. The relevant nuclear matrix elements (NMEs) involved in the process are calculated within the formalism of the microscopic interacting boson model (IBM-2).
The IBM-2 has been widely used to obtain predictions for nuclear observables, such as the spectrum, but also to explore the possible emergence of beyond-the-Standard Model effects in the weak interactions of nuclei.
Our calculations are carried out by considering the exchange of a Majorana neutrino between two nucleons ($2N$-mechanism). In addition to NMEs, we calculate the associated  leptonic phase-space factors (PSFs) using electron radial wave functions, which are obtained by solving numerically the Dirac equation of a screened Coulomb potential that takes into account finite nuclear size.
By combining our IBM-2 results for the NMEs with those for the PSFs along with experimental half-life limits, we can set limits on the $\langle \lambda \rangle$ and $\langle \eta \rangle$ couplings of left-right (L-R) models.
\end{abstract}

\maketitle

\section{INTRODUCTION}
\label{sec:introd}
Neutrinos have a long story. Their existence was postulated by Pauli in 1930 to ensure the conservation of energy and angular momentum in $\beta$-decay \cite{Pauli:1930pc}. Fermi's renowned theory of beta decay dates back to 1933 \cite{Fermi:1934hr}. In 1956, neutrinos were first observed at Los Alamos by Cowan and Reines via the study of inverse beta decay \cite{Cowan:1992xc}.
Several decades later, neutrinos are still fascinating and mysterious particles. 

Important questions regarding some of their main properties remain unsolved, including the unknown mechanism that generates their masses and a complete understanding of their mixing mechanism and mass hierarchy \cite{Giunti:2003qt}.
Because of the lack in the standard model (SM) of a Yukawa coupling between the Higgs boson and neutrinos, due to the absence of right-handed neutrinos, the SM has to be extended to provide a neutrino mass term. Extensions of the SM include the L-R symmetric \cite{Pati:1974yy,Mohapatra:1974hk,Senjanovic:1975rk} and SUSY \cite{Mohapatra:1986su,Vergados:1986td,Hirsch:1995zi,Babu:1995vh,Hirsch:1995cg,Faessler:1996ph} models.

Some important issues are directly related to the nature of neutrinos as Fermi- or Majorana-type particles, a nature which could be directly assessed via the experimental observation of neutrinoless double-beta ($0\nu\beta\beta$) decay process \cite{DellOro:2016tmg}.
However, despite of the strenuous attempts by many experimental groups, e.g.,  \cite{Gando:2012zm,Agostini:2013mzu,Alfonso:2015wka,Auger:2012ar,Arnold:2015wpy,Aalseth:2017btx,Azzolini:2019yib}, $0\nu\beta\beta$-decay has  not yet  been observed.

Several theoretical investigations on $0\nu\beta\beta$ decay have been published over the years (for a review see e.g., Refs. \cite{ejiri2019,Agostini:2022zub}) in order to guide the experimentalists in their searches.
Owing to the low-energy character of $0\nu\beta\beta$ processes, these studies necessarily involve elements of both particle and nuclear physics.
In particular, nuclear structure models are necessary in order to take care of the nuclear matrix elements (NMEs) \cite{Barea:2013bz,Simkovic:2007vu,Caurier:2007wq} entering the expression of the $0\nu\beta\beta$-decay half-life. 

Here, we show the results of a calculation of the $0^+$ to $2^+$ $0\nu\beta\beta$ decay of $^{76}$Ge, $^{82}$Se, $^{130}$Te and $^{136}$Xe, in which we consider the exchange of a Majorana neutrino between two nucleons, the so-called $2N$-mechanism, within an L-R symmetric model \cite{Doi:1985dx,Tomoda:1990rs}.
Specifically, in our study: I) we compute the relevant NMEs within the microscopic interacting boson model (IBM-2) formalism \cite{Iachello:2006fqa} and compare our results with previous calculations for the studied nuclei within different nuclear structure models; II) we calculate the leptonic phase-space factors (PSFs) by means of electron radial wave functions, obtained by solving numerically the Dirac equation of a screened Coulomb potential that takes into account finite nuclear size \cite{Kotila:2012zza}; and III) by combining the two above elements, namely the results for the NMEs and the leptonic PSFs, with the experimental limits on the half-life, we set limits on the $\langle \lambda \rangle$ and $\langle \eta \rangle$ couplings of L-R models. Experimental studies on this decay mode can be found e.g., in \cite{HM,PhysRevC.103.015501,LUCIFER:2015ozh,CUPID:arxiv,Arnaboldi:2002te,KamLAND-Zen:2015tnh}. Previous calculations for the $0^{+} \rightarrow 2^{+}$ decay rate for the studied nuclei within different nuclear structure models and via the $2N$-mechanism can be found in Refs. \cite{Tomoda:1990rs,Doi:1985dx,Tomoda:1999zc,Fang:2021tfc,Tomoda:1988ew}.

This article is organized as follows: In Sec. \ref{0+ 2+ LR} the importance of $0^+ \rightarrow 2^+$ $0\nu\beta\beta$ decay is discussed and some details on the calculation of $0\nu\beta\beta$ decay rates in L-R symmetric models are provided. In Sec. \ref{nuclear wave functions} the calculation of IBM-2 wave functions is briefly summarized, and  in Sec. \ref{sec3} the decay operators needed for the description of $0^+ \rightarrow 2^+$ $0\nu\beta\beta$-decay are presented. In Sec. \ref{decay rate and PSF} the numerical results for the ingredients needed for the calculation of the decay rate are given and discussed. Finally, the conclusions are presented in Sec. \ref{sec:Conc}.

\section{$0^+ \rightarrow 2^+$ $0\nu\beta\beta$ decay in the L-R symmetric model}
\label{0+ 2+ LR}
If we restrict ourselves only to long-range mechanisms for $0\nu\beta\beta$ decay, the most general effective Lagrangian is the Lorenz-invariant combination of leptonic, $j_\alpha$, and hadronic, $J_\alpha$, currents with definite tensor structure and chirality \cite{Kotila:2021xgw,Ali:2007ec,Pas:1999fc},
\begin{equation}
	\label{eqn:effective-L}
	{\mathcal L} = \frac{G_{\rm F} \cos \theta_{\rm c}}{\sqrt 2} \left[ U_{ei} ~ j_{V-A}^{\mu,i} J_{V-A,\mu}^\dag
	+ \displaystyle \sum_{\alpha,\beta} \epsilon_{\alpha,i}^\beta ~ j_\beta^i J_\alpha^\dag + {\rm h.c.} \right] \mbox{ }.
\end{equation}
Here, $G_{\rm F}$ is the Fermi constant; $\theta_{\rm c}$ is the Cabibbo angle; the hadronic and leptonic currents are defined as $J_\alpha^\dag = \bar u \mathcal O_\alpha d$ and $j_\beta^i = \bar e \mathcal O_\beta \nu_i$, where the index $i$ spans the neutrino mass eigenstates. The indices $\alpha, \beta$ are $V\mp A$, $S\mp P$, $T\mp T_5$, where $S$, $P$, $T$ and $T_5$ stand for scalar, pseudo-scalar, tensor, and pseudo-tensor, respectively.
In Eq. (\ref{eqn:effective-L}) we have isolated the standard model contribution proportional to $U_{ei}$, where $U_{ei}$ is the PMNS mixing matrix element \cite{Pontecorvo:1957qd,Maki:1962mu}, from non-standard contributions, which are those proportional to the couplings $\epsilon_{\alpha,i}^\beta$.

By isolating the $V\mp A$ currents of the L-R models from those allowed in other types of mechanisms, such as SUSY, and performing a non-relativistic reduction of both the leptonic and hadronic currents, one can obtain the expression of the $0\nu\beta\beta$ decay half-life for a $0^+ \rightarrow 0^+$ transition in L-R models:
\begin{equation}
	\label{eqn:rate-LR}
	\begin{array}{lll}
	\left[ \tau_{1/2}^{0\nu} (0^+ \rightarrow 0^+) \right]^{-1} &=& C_{mm}^{(0)}\left( \frac{\left\langle m_{\nu }\right\rangle }{%
m_{e}}\right) ^{2}+C_{\lambda \lambda }^{(0)}\left\langle \lambda\right\rangle ^{2}\\
&+&C_{\eta \eta }^{(0)}\left\langle \eta \right\rangle^{2}+2C_{m\lambda }^{(0)}\frac{\left\langle m_{\nu }\right\rangle }{m_{e}}%
\left\langle \lambda \right\rangle  \\
&+&2C_{m\eta }^{(0)}\frac{\left\langle m_{\nu }\right\rangle }{m_{e}}%
\left\langle \eta \right\rangle +2C_{\lambda \eta }^{(0)}\left\langle
\lambda \right\rangle \left\langle \eta \right\rangle.
\end{array}
\end{equation}
The above equation is a complicated combination of the three parameters $\langle m_\nu \rangle$, $\langle \lambda \rangle$, and $\langle \eta \rangle$ and their respective matrix elements and phase-space factors; for details on the combinations $C_{ij}^{(0)}, i,j=m, \lambda, \eta$ see e.g., Ref. \cite{Kotila:2021xgw}.

The new physics beyond the standard model is enclosed in the three parameters
\begin{subequations}
\begin{equation}
	\label{eqn:M-nu}
	\langle m_\nu \rangle =\sum_i U_{ei}^2 m_i \mbox{ },
\end{equation}
\label{eqns:lambda-eta}
\begin{equation}
	\langle \lambda \rangle= \lambda \sum_{j} U_{ej}V_{ej} \mbox{ },
\end{equation}
and
\begin{equation}
	\langle \eta \rangle= \eta \sum_{j} U_{ej}V_{ej} \mbox{ },
\end{equation}
\end{subequations}
where  $\langle m_\nu \rangle$ is the average neutrino mass obtained by summing over mass $m_i$ of neutrino species $i$, and $\langle \lambda \rangle$ and $\langle \eta \rangle$ are the standard couplings of L-R models \cite{Tomoda:1990rs}. $U_{ej}$ and $V_{ej}$ are the mixing matrix elements of the PMNS matrix \cite{Pontecorvo:1957qd,Maki:1962mu} for the standard and non-standard (L-R) mechanisms, respectively. Thus, the $0^+ \rightarrow 0^+$ process can occur because either of right- or left-handed leptonic currents.

By contrast, if  $0\nu\beta\beta$-decay to a $2^+$ state is observed then in addition to proving the Majorana character of neutrinos,  the existence of  $V+A$ current would also be established, since as a first approximation, this decay mode is  triggered by right-handed leptonic currents only \cite{Tomoda:1988ew}. 
To be more specific, the combination of the lowest electron partial waves for the  $0^+ \rightarrow 2^+$  transition is
the $S$ and  $P_{3/2}$ case since the total angular momentum of the two-electron system should be 2. However, in order to have a non-zero contribution of the L-L term that is proportional to the average neutrino mass, $\langle m_\nu \rangle$, the leading term requires the combination of $S$ and $D_{3/2}$ electron waves,  making it negligible compared with the contributions due to L-R terms,  which are those proportional to $\langle \lambda \rangle$ and $\langle \eta \rangle$). For more details see \cite[App. C]{Doi:1985dx}.

Therefore, for the $0^+ \rightarrow 2^+$ case the half-life can safely be written without the dependence on the average neutrino mass 
\begin{equation}
	\label{eqn:tau-lam-eta}
	\begin{array}{lll}	
	\left[\tau_{1/2}^{0\nu\beta\beta}(0^+ \rightarrow 2^+)\right]^{-1} 
 	&=&  g_{\rm A}^4  \Big[ G_{1} \left(M_{\lambda} 
	\langle \lambda \rangle \right.  
	 - \left. M_{\eta} \langle \eta \rangle \right)^2 \\
	&+& G_{2}\left(M'_{\eta} \langle \eta \rangle\right)^{2}  \Big]
	\end{array} \mbox{ },
\end{equation}
where one can factorize the leptonic phase-space factor (PSF) \cite{Tomoda:1990rs,Kotila:2012zza}, $G$, the nuclear matrix elements (NMEs), $M$, and the axial vector coupling constant, $g_{\rm A}$. Neutrinoless double-beta decay to a  $2^+$ state thus provides information that is different from that one could gather from the study of $0^+ \rightarrow 0^+$ $0\nu\beta\beta$ processes. Moreover, the observation of a $0^+ \rightarrow 2^+$ $0\nu\beta\beta$-decay may also  possibly rule out those non-standard mechanisms in which no right-handed gauge bosons or fermions are present. 

\begin{table*}[tb!]
\begin{ruledtabular}
\begin{centering}
\begin{tabular}{lccccccccccccc}
Nucleus  & $\epsilon_{d_{\nu}}$  & $\epsilon_{d_{\pi}}$  & $\kappa$  & $\chi_{\nu}$  & $\chi_{\pi}$  & $\xi_{1}$  & $\xi_{2}$  & $\xi_{3}$  & $c_{\nu}^{(0)}$  & $c_{\nu}^{(2)}$    & $c_{\pi}^{(0)}$  & $c_{\pi}^{(2)}$  & $c_{\pi}^{(4)}$  \tabularnewline
\hline 
$^{76}\mbox{Ge}$ \cite{Duval:1983tit}  & 1.20  & 1.20  & -0.21  & 1.00  & -1.20  & -0.05  & 0.10  & -0.05  &  &  &  &  &    \tabularnewline
$^{76}\mbox{Se}$ \cite{Kaup83}  & 0.96  & 0.96  & -0.16  & 0.50  & -0.90  &  &  & -0.10  &  &  &  &  &  \tabularnewline
$^{82}\mbox{Se}$ \cite{Kaup83}  & 1.00  & 1.00  & -0.28  & 1.14  & -0.90  &  &  & -0.10  &  &  &  &  &     \tabularnewline
$^{82}\mbox{Kr}$ \cite{Kaup79}  & 1.15  & 1.15  & -0.19  & 0.93  & -1.13  & -0.10  &  & -0.10  &  &  &    &  &  \tabularnewline
$^{130}\mbox{Te}$ \cite{Sambataro82}  & 1.05  & 1.05  & -0.20  & 0.90  & -1.20  & -0.18  & 0.24  & -0.18  & 0.30  & 0.22  &  &    &   \tabularnewline
$^{130}\mbox{Xe}$ \cite{Puddu80}  & 0.76  & 0.76  & -0.19  & 0.50  & -0.80  & -0.18  & 0.24  & -0.18  & 0.30  & 0.22    &  &    & \tabularnewline
$^{136}\mbox{Xe}$\footnotemark[1]  &  & 1.31  &  &  &  &  &  &  &  &    & -0.04  & 0.01  & -0.02   \tabularnewline
$^{136}\mbox{Ba}$ \cite{Puddu80}  & 1.03  & 1.03  & -0.23  & 1.00  & -0.90  & -0.18  & 0.24  & -0.18  & 0.30  & 0.10  &  &  &    \tabularnewline
\end{tabular}
\footnotetext[1]{GS parameters fitted to reproduce the spectroscopic data of the low-lying energy states.}
\par\end{centering}
 \end{ruledtabular} 
\caption{Hamiltonian parameters employed in the IBM-2 calculation, along with their references. All the values are in MeV, with the exception of those of $\chi_\pi$ and $\chi_\nu$, which are dimensionless. The IBM-2 parameters not shown here are set to zero.}
\label{tab:IBM-2-Model-parameters}
\end{table*}

\section{IBM-2 nuclear wave functions}
\label{nuclear wave functions}
IBM-2 is a nuclear structure model and was originally introduced as a phenomenological approach to describing collective excitations in nuclei \cite{Arima:1976ky}.
Soon afterwards, its relation with the shell model was established \cite{Arima:1977vie,Otsuka:1978zza,Otsuka:1978zz}. The starting point of IBM-2 calculations of any nuclear observable, including weak  decay rates, is to obtain the nuclear wave functions of the nuclei of interest.
Realistic nuclear wave functions are obtained by fitting the IBM-2 parameters in order to reproduce the experimental energy levels and other nuclear properties, such as electromagnetic transition rates, quadrupole, and magnetic moments etc. \cite{Iachello:2006fqa,NPBOS, Duval:1983tit,Kaup79,Kaup83,Sambataro82,Puddu80}, and the relevant two-body operators are derived in the IBM-2 formalism \cite{Arima:1977vie,Otsuka:1978zz,Barea:2009zza}.

The IBM-2 Hamiltonian describing the spectra of even-even nuclei, which is often used in literature, and which is general enough for the phenomenological studies, reads \cite{Otsuka:1978zz,Iachello:2006fqa}
\begin{equation}	
	\label{eqn:H-IBM}
	\begin{array}{rcl}
	H_{\rm B} & = & \epsilon_d (\hat n_{d_\pi} + \hat n_{d_\nu}) - \kappa \left(Q_\nu^{\rm B}\cdot Q_\pi^{\rm B}\right) 	\\
	& + & \frac{1}{2} \xi_2  \left[\left(d_\nu^\dag s_\pi^\dag - d_\pi^\dag s_\nu^\dag\right) \cdot 
	\left(\tilde d_\nu s_\pi - \tilde d_\pi s_\nu\right) \right] \\
	& + & \displaystyle \frac{1}{2}
\sum_{\rho}
\sum_{K=1,3} \xi_K 
	\left[d_\nu^\dag \times d_\pi^\dag\right]^{(K)} \cdot \left[\tilde d_\pi \times \tilde d_\nu\right]^{(K)}\\
	& + & \displaystyle \sum_{L=0,2,4}C^{\rho}_{L}
\Big(
[d^{\dagger}_{\rho}d^{\dagger}_{\rho}]^{(L)}\cdot 
[\tilde d_{\rho}\tilde d_{\rho}]^{(L)}
\Big) \mbox{ }.
	\end{array} 
\end{equation}
\begin{table*}[htbp]
\centering
\begin{ruledtabular}
\begin{tabular}{cccccc}
i & $M_{i}$ & $C_{\lambda i}$ & $C_{\eta i}$ & $C'_{\eta i}$ \\
\hline
1 & ${\bm \sigma}_{1} \cdot {\bm \sigma}_{2}[\hat {\bf r}_{12} \times \hat {\bf r}_{12}]^{(2)}h(r_{12})$ & $\frac{1}{3}$ & $\frac{1}{3}$ & -- \\
2 & $[{\bm \sigma}_{1}\otimes {\bm \sigma}_{2}]^{(2)}h(r_{12})$ & $-\frac{2}{3}$ & $-\frac{2}{3}$ & -- \\
3 & $\left[[{\bm \sigma}_{1} \times {\bm \sigma}_{2}]^{(2)} \times [\hat {\bf r}_{12}\otimes \hat {\bf r}_{12}]^{(2)}\right]^{(2)}h(r_{12})$ & $\sqrt{\frac{7}{3}}$ & $\sqrt{\frac{7}{3}}$ & -- \\
4 & $[\hat {\bf r}_{12} \times \hat {\bf r}_{12}]^{(2)}h(r_{12})$ & $(g_{\rm V}/g_{\rm A})^{2}$ & $-(g_{\rm V}/g_{\rm A})^{2}$ & -- \\
5 & $\left[({\bm \sigma}_{1}+{\bm \sigma}_{2}) \times [\hat {\bf r}_{12} \times \hat {\bf r}_{12}]^{(2)}\right]^{(2)}h(r_{12})$ & $-\sqrt{\frac{3}{2}} (g_{\rm V}/g_{\rm A})$ & -- & -- \\
6 & $\left[({\bm \sigma}_{1}-{\bm \sigma}_{2}) \times [\hat {\bf r}_{12} \times \hat {\bf r}_{+12}]^{(1)}\right]^{(2)} \frac{r_{+12}}{r_{12}} h(r_{12})$ & -- & -- & $\sqrt{\frac{1}{2}}(g_{\rm V}/g_{\rm A})$ \\
7 & $\left[({\bm \sigma}_{1}-{\bm \sigma}_{2}) \times [\hat {\bf r}_{12} \times \hat  {\bf r}_{+12}]^{(2)}\right]^{(2)} \frac{r_{+12}}{r_{12}} h(r_{12})$ & -- & -- & $-\sqrt{\frac{3}{2}}(g_{\rm V}/g_{\rm A})$ \\
\end{tabular}
\end{ruledtabular}
\caption{0$^{+}$ $\rightarrow$ 2$^{+}$ $ 0\nu\beta\beta$ decay via the $2N$-mechanism. Here, we enlist the two-body operators, $M_{i}$, giving the dominant contribution, as well as the coefficients $C_{\lambda i}$, $C_{\eta i}$ and $C'_{\eta i}$ of the two-body operators \cite{Tomoda:1988ew}. The coefficients which are not given explicitly are null.}
\label{cof}
\end{table*}
In the previous expression, $\hat n _{d_\rho} = d_\rho^\dag d_\rho$ and 
\begin{equation}
	\label{eqn:Q-rho}
	Q_\rho^{\rm B} = d_\rho^\dag s_\rho + s_\rho^\dag \tilde d_\rho + \chi_\rho [d_\rho^\dag \times\tilde d_\rho]^{(2)}
\end{equation}
represent the $d$-boson number operators and the boson quadrupole operators for the proton ($\rho = \pi$) and neutron ($\rho = \nu$) pairs, respectively; $s_\rho^\dag$ and $d_\rho^\dag$ are $s_\rho$- and $d_\rho$-boson creation operators, and the modified $d_\rho$-boson annihilation operator satisfies $\tilde d_{\rho,m} = (-1)^m d_{\rho,-m}$. The third term on RHS of Eq.~(\ref{eqn:H-IBM}) is the so-called Majorana term, which is relevant to the proton-neutron mixed symmetry, and has been considered, e.g., in the context of the isovector
collective motion of valence shells. The last term on the RHS of Eq.~(\ref{eqn:H-IBM}) corresponds to the interaction between like bosons, and consists of $L=0, 2$ and 4 components, respectively. 

A detailed description of the IBM-2 Hamiltonian is given in Refs. \cite{Arima:1977vie} and \cite{NPBOS}. The Hamiltonian parameters are taken from the literature \cite{Duval:1983tit,Kaup83,Kaup79,Sambataro82,Puddu80,Barea:2013bz}. The values of the Hamiltonian parameters, together with the references from which they are taken, are reported in Table~\ref{tab:IBM-2-Model-parameters}.

\section{0$\nu\beta \beta$ decay operators for 0$^{+} \rightarrow 2^{+}$ transitions in the IBM-2}
\label{sec3}
In the present study, we focus on the $2N$-mechanism discussed by Tomoda \cite{Tomoda:1988ew} in the context of L-R models \cite{Pati:1974yy,Mohapatra:1974hk,Senjanovic:1975rk,Hirsch:1996qw,Doi:1985dx}. 
To do so, we need to consider a specific set of $V^{(2)}_{i} = M_{i} C_{\kappa i}$ operators, where the index $\kappa$ refers to the $\lambda$, $\eta$ and $\eta'$ contributions of L-R models.
The coefficients $C_{\kappa i}$ and the corresponding two-body operators $M_i$ are enlisted in Table \ref{cof}.

The seven operators in this table can be written as a combination of three parts: the relative coordinate, $O_i (\hat {\bf r}_{12})h(r_{12})$, the center-of-mass coordinate, $O_i (\hat {\bf r}_{+12}) f (r_{+12})$, and the spin part, $O_i ({\bm \sigma}_{1}, {\bm \sigma}_{2} )$,  where the following notation for the coordinates is used: ${\bf r}_{12} = {\bf r}_{2} - {\bf  r}_{1}$, ${\bf  r}_{+12} = {\bf  r}_{2}+ {\bf  r}_{1}$, $\hat {\bf r} = {\bf  r} / |{\bf  r}|$. 

The neutrino potential, which comes from the electron-Majorana neutrino exchange, is given by \cite{Tomoda:1986yz,Vergados:1988xp}
\begin{equation}
	\label{eqn:h}
	h(r_{12}) = -r_{12} \frac{\partial}{\partial r_{12}} \, H(r_{12},\overline A)  \mbox{ }.
\end{equation}
In the above equation,
\begin{equation}
	\overline A= \langle E_{N}\rangle - E(0^{+})+m_{e}+\frac{1}{2}Q_{\beta \beta}(2^{+})  
\end{equation}
is the closure energy, where $ E(0^{+})$ is the energy of the initial $0^{+}$ state, $\langle E_{N}\rangle $ the average energy of the intermediate excited state, and $Q_{\beta \beta}(2^{+})$ is the $Q$-value of the $0^{+} \rightarrow 2^{+}$  decay process. The neutrino propagation function in Eq. (\ref{eqn:h}), $H(r_{12},\overline A)$, is given by \cite{Tomoda:1986yz,Vergados:1988xp}
\begin{equation}
	\label{eqn:H-r12}
	H(r_{12},\overline A)=\frac{4\pi}{(2\pi)^{3}}\int d{\bf p}_{12} \frac{\text{exp}(i{\bf p}_{12}\cdot {\bf r}_{12})}{p_{12}
	(p_{12}+\overline A)}  \mbox{ },
\end{equation}
where ${\bf p}_{12}$ is the conjugate momentum to the ${\bf r}_{12}$ coordinate. More details on the neutrino potential can be found in e.g., \cite{Barea:2009zza}.

On introducing a proton (neutron) creation (annihilation) operator $\pi^\dagger_{nljm}(\tilde{\nu}_{nljm})$ that acts on the single-particle state $|nljm\rangle$, the second quantized fermion operator  can be written as
\begin{widetext}
\begin{equation}  
\label{Formula1}
\begin{array}{lll}
	V^{(\lambda)}_i & = & \displaystyle  -\frac{1}{4}  \sum_{j_{1},j_{2}} \sum_{ j_{1'},j_{2'}}\sum_{ J,J'} (-1)^{J+J'}\sqrt{1+(-1)^{J}\delta_{j_{1},j_{2}}} 
	\sqrt{1+(-1)^{J'}\delta_{j_{1'},j_{2'}}} \\
	& & \times {\mathcal G}_{i}(j_{1},j_{2},J,j_{1'},j_{2'},J',\lambda)(\pi^\dagger_{j_1}\times \pi^\dagger_{j_2})^{(J)}(\tilde{\nu}_{j'_1}\times \tilde{\nu}_{j'_2})^{(J')} \mbox{ },
\end{array} 
\end{equation}
with $J,J'=0,2$ for the current study, and for $i=1-5$
\begin{equation}
	\begin{array}{lll}        
	{\mathcal G}_{i} & = & \sqrt{\frac{2}{3}}\displaystyle \sum_{k k'} \sum_{k_{1}k_{2}}i^{k_{1}-k_{2}+\lambda_{2}} 
	\frac{\hat{k}_{1}^2\hat{k}_{2}^2}{\hat{\lambda}_{2}^2} \, \langle k_{1} 0 k_{2} 0 \lambda_{2}0 \rangle \, v^{k_{1},k_{2};
	\lambda_{2}}(r_{1}, r_{2}) \, \hat k \hat k'  \hat \lambda_{1} \hat \lambda_{2} 
	\left\{\begin{array}{cccc} s_1 & k_1 & k \\ s_2 & k_2 & k' \\ \lambda_1 & \lambda_2 & 
	\lambda \end{array}\right\}  \hat J \hat \lambda \hat J' \left\{\begin{array}{cccc} j_1 & j_2 & J \\ j_1' & j_2' &J' \\
	k & k' & \lambda \end{array} \right\} \\
	& &\times 
	\hat j_{1} \hat k \hat j_{1}' \left\{\begin{array}{cccc} \frac{1}{2} & l_1 & j_1 \\
	\frac{1}{2} & l_1' & j_1' \\ s_1 & k_1 & k \end{array}\right\} 
	\hat j_{2} \hat k' \hat j_{2}' \left\{\begin{array}{cccc} \frac{1}{2} & l_2 & j_2 \\ \frac{1}{2} & l_2' & j_2' \\ 
	s_2 & k_2 & k \end{array}\right\}
	\langle \frac{1}{2}\| \Sigma^{(s_1)}\|\frac{1}{2}\rangle 
	\, (-1)^{k_1} \hat l_{1}  \langle l_{1} 0 k_{1} l_{1}' 0 \rangle    \langle \frac{1}{2}\| \Sigma^{(s_{2})}\|\frac{1}{2}\rangle (-1)^{-k_{2}}  \\ 
	& & \times  \hat l_{2}  \langle l_{2} 0 k_{2} l_{2}' 0 \rangle
	R^{(k_{1}k_{2}\lambda_{2})}(n_{1},l_{1},n_{2},l_{2},n_{1}',l_{1}',n_{2}',l_{2}') \mbox{ }.
\end{array}  
\label{twobody2+}
\end{equation}
\end{widetext}
In Eq. (\ref{twobody2+}), $\Sigma^{(1)} = {\bm \sigma}$, $\Sigma^{0}=1$, $\langle \frac{1}{2}\| \Sigma^{(s)}\|\frac{1}{2}\rangle=\sqrt{2(2s+1)}$, $|l_i-l'_i| \leq k_i\leq l_i+l'_i$, $|j_1-j'_1| \leq k\leq j_1+j'_1$, and $|j_2-j'_2| \leq k'\leq j_2+j'_2$. Also, an additional factor of $-\sqrt{3}$ is needed for ${\mathcal G}_{1}$ and of $-\sqrt{3/2}$ for ${\mathcal G}_{2}$. ${\mathcal G}_{5}$ is evaluated in two parts by applying additional factor of $1/3\sqrt{1/6}$.  In our current study the protons and neutrons occupy the same major shell and thus the contributions of ${\mathcal G}_6$ and ${\mathcal G}_7$ vanish. Regarding the radial integrals, indicated as $R^{(k_{1}k_{2}\lambda_{2})}(n_{1},l_{1},n_{2},l_{2},n_{1}',l_{1}',n_{2}',l_{2}')$ in Eq. (\ref{twobody2+}), their evaluation follows the procedure discussed in \cite{Horie1961}.

 The calculation of the nuclear matrix elements of $0\nu\beta\beta$ decay would, in principle, proceed by going through all the virtual intermediate states in the odd-odd nucleus. However, this is a demanding task, which can be greatly simplified by treating the sum over the intermediate states in the closure limit. This is a good approximation in the case of $0\nu\beta\beta$ decay, as the energy of the virtual neutrino exchange between nucleons is much larger than the typical excitation energy of the intermediate states \cite{Doi:1985dx,Tomoda:1990rs}. In the closure approximation, one is left with the calculation of two-body NMEs between even-even initial and final states.

In order to obtain the bosonic image of the fermionic $0\nu\beta\beta$ decay operator, we proceed in a similar way to Ref. \cite{Barea:2009zza}. The following step is then to define a mapping between the IBM-2 $J^P = 0^+$ and $2^+$ boson creation operators, $s^\dag$ and $d^\dag$, and the shell-model creation operators of collective nucleon pairs, $S^\dag = \left( \rho_j^\dag \times \rho_j^\dag \right)^{(0)}$ and $D^\dag = \left( \rho_j^\dag \times \rho_{j'}^\dag \right)^{(2)}_M$, where the fermion operators $\rho^\dag_j$ create nucleons (either neutrons, $\rho = \nu$, or protons, $\rho = \pi$) with angular momentum $j$ \cite{Arima:1977vie,Otsuka:1978zz,Barea:2009zza}.  
This procedure enables us to find a direct correspondence between the matrix elements between fermionic states in the $SD$ shell-model subspace and the matrix elements in the $sd$ bosonic space of the IBM-2.
One has
\begin{subequations}
\label{mapping}
\begin{equation}
	\left( \rho_j^\dag \times \rho_j^\dag \right)^{(0)} \rightarrow  A_{\rho}(j) \, s_{\rho}^{\dagger}
\end{equation}
and
\begin{equation}
	\left( \rho_j^\dag \times \rho_{j'}^\dag \right)^{(2)}_M \rightarrow B_{\rho}(j,j') \, d_{\rho,M}^{\dagger} \mbox{ },
\end{equation}
\end{subequations}
where the mapping coefficients $A_{\rho}(j)$ and $B_{\rho}(j,j')$ are obtained by means of the OAI method \cite{Otsuka:1978zz} and depend on the specific normalization of the nuclear structure coefficients that one considers. Our choice is to use the conventions for the $A_{\rho}(j)$ and $B_{\rho}(j,j')$ coefficients reported in Refs. \cite{Mardones:2016wgy,Barea:2014lza}, which are based on the procedure for diagonalizing the Surface Delta Interaction (SDI) of Ref. \cite{Pittel:1982} and the use of the commutator method of Refs. \cite{Frank:1982zz,Lipas:1990rs,Barea:2009zza}. 

\section{0$_{1}^{+}$ to 2$_{1}^{+}$ $0\nu\beta\beta$ decay}
\label{decay rate and PSF}

\subsection{Nuclear matrix elements}
Here, we give results for the NMEs relevant to the 0$_{1}^{+}$ to 2$_{1}^{+}$ $0\nu\beta\beta$ decay processes of $^{76}$Ge, $^{82}$Se, $^{130}$Te and $^{136}$Xe.
The nuclear matrix elements of the operators $M_i$ (with $i = 1, ...,7$) in Table \ref{cof} are computed in the IBM-2 formalism \cite{Arima:1977vie,Otsuka:1978zz,Barea:2009zza}. 
The calculations can be made more realistic by introducing short-range correlation effects, i.e. the two-body operators in Table \ref{cof}  need to be multiplied by the short-range correlation function, $f(r)$, squared. Following Ref. \cite{Barea:2009zza}, we make use of the Jastrow function,
\begin{equation}
	f(r) = 1 - c e^{-a r^2} (1 - b r^2)  \mbox{ },
\end{equation}
with Argonne parametrization $a = 1.59$ fm$^{-2}$, $b = 1.45$ fm$^{-2}$ and $c = 0.92$ \cite{argonne}.	

The finite size of the nucleon is taken into into account by substituting the coupling constants $g_{\rm A}$ and $g_{\rm V}$ with the form factors,
\begin{subequations}
\label{eqns:FFs}
\begin{equation}
	g_{\rm V}(p_{12}^2) = \frac{g_{\rm V}}{\left(1+\frac{p_{12}^2}{M_{\rm V}^2}\right)^2}  \mbox{ }
\end{equation}
and
\begin{equation}
	g_{\rm A}(p_{12}^2) = \frac{g_{\rm A}}{\left(1+\frac{p_{12}^2}{M_{\rm A}^2}\right)^2}  \mbox{ }.
\end{equation}
\end{subequations}
In the above equations, the constant $M_{\rm V}^2 = 0.71$ GeV$^2$ is fixed by the electromagnetic form factor of the nucleon \cite{Iachello:1972nu,Bijker:2004yu} and the value of $g_{\rm V} = 1$ by the conserved vector current (CVC) hypothesis; the value of $M_{\rm A}^2 = 1.09$ GeV$^2$ is determined from neutrino scattering data \cite{Schindler:2006jq} and that of $g_{\rm A}$ from neutron decay \cite{Zyla:2020zbs}.


\begin{table*}[htbp]
\centering
\begin{ruledtabular}
\begin{tabular}{ccccccccccc}
 & $M_{1}$ & $M_{2}$ & $M_{3}$ &$M_{4}$ & $M_{5}$ & $M_{6}$ & $M_{7}$ &$M_{\lambda }$ & $M_{\eta}$ & $M'_{\eta}$ \\
\hline
$^{76}$Ge &0.189 &-0.056 &-0.023 &-0.069 &-0.013 &0 &0 &0.035&0.108&0\\ 
$^{82}$Se &0.003 &0.080 &-0.003&-0.007&-0.011&0&0&-0.051&-0.053&0 \\
$^{130}$Te &0.153 &-0.081 &-0.016&-0.050&-0.006&0&0&0.056&0.112&0 \\
$^{136}$Xe &0.058 &-0.112 &-0.006&-0.012&-0.0001&0&0&0.077&0.092&0 \\
\end{tabular}
\end{ruledtabular}
\caption{Nuclear matrix elements for 0$^{+}$ $\rightarrow$ 2$^{+}$ $ 0\nu\beta\beta$ decay via the $2N$-mechanism obtained by using IBM-2.}
\label{num-NMEs}
\end{table*}
Our  IBM-2 results  for $^{76}$Ge, $^{82}$Se, $^{130}$Te, and $^{136}$Xe are given in Table \ref{num-NMEs}. In the last three columns, we give the combined NMEs 
\begin{equation}
	\begin{array}{l}
	M_{\lambda}=\displaystyle \sum_{i=1}^{5}C_{\lambda i}M_{i} \,, \mbox{ } \,
	M_{\eta}=\displaystyle \sum_{i=1}^{5}C_{\eta i}M_{i} \,, \\
	M'_{\eta}=\displaystyle  \sum_{i=6}^{7}C'_{\eta i}M_{i} \,,
	\end{array}
\end{equation}
where the values of the coefficients $C_{\lambda i}$, $C_{\eta i}$ and $C'_{\eta i}$ are given in Table \ref{cof}. The values in Table \ref{num-NMEs} are calculated by using unquenched values of $g_V=1$ and $g_A=1.269$, and quenching can be implemented through coefficients $C$. 

It is worth noting that, in the nuclei considered here, protons and neutrons occupy the same major shell and thus the contributions of $M_6$ and $M_7$ vanish, leading to $M'_{\eta}=0$. One also notices particularly small $M_1$ value for the case of $^{82}$Se  originating from the small bosonic $d^\dagger_\pi \tilde{d}_\nu$ matrix element. The same also happens for the first excited $0^+$ state in $^{82}$Se decay, leading to small IBM-2 $^{82}\mbox{Se}(0^+_1)  \rightarrow{}^{82}\mbox{Kr}(0^+_2)$ $0\nu\beta\beta$ NME, as can be seen from Ref. \cite{Barea:2015kwa}.

Our results can be compared to those of the few existing studies on this topic. Specifically, the $0^+ \rightarrow 2^+$ $0\nu\beta\beta$ decay of $^{76}{\rm Ge}$ was studied in Refs. \cite{Tomoda:1988ew} and \cite{Fang:2021tfc} by means of projected the Hartree Fock Bogoliubov (PHFB) method and quasiparticle random-phase approximation (QRPA), respectively. We observe that the first operator provides the largest contribution to both $M_\lambda$ and $M_\eta$ in all of these three calculations. Our results, $M_\lambda=0.035$ and $M_\eta=0.108$, stand in between the PHFB results ($M_\lambda=0.002$ and $M_\eta=0.061$) and QRPA results ($M_\lambda=0.008-0.228$ and $M_\eta=0.317-0.540$).


\subsection{Leptonic phase-space integrals}
The leptonic phase-space integrals, indicated as $G_i$ $(i=1,2)$ in Eq. (\ref{eqn:tau-lam-eta}), are given by \cite{Tomoda:1988ew,Tomoda:1990rs}
\begin{equation}
\label{eqn:Gi}
G_i=\frac{2}{\ln2}\frac{(G_{\rm F}\cos\theta_{\rm C})^4}{16\pi^5}\frac{m_e^2}{4R^2_A}\int^{Q_{\beta\beta}^{2^+}+m_e^2}_{m_e^2}f_i p_1p_2E_1E_2dE_1,
\end{equation}
with $E_2=Q_{\beta\beta}^{2^+}+m_e^2-E_1$.
In the above equation, $G_{\rm F} = 1.1663787 (6) \cdot 10^{-5}$ GeV$^{-2}$ is the Fermi coupling constant and $\theta_{\rm C}$ the Cabibbo angle \cite{Zyla:2020zbs}; 
$E_{j}$ and $p_{j}=\sqrt{E_{j}^{2}-m_{e}^{2}}$ are the energies and the asymptotic momenta of the electrons, respectively;
\begin{subequations}
\begin{equation}
	\begin{array}{rcl}
	f_{1} & = & \frac{3}{(m_{e}R_A)^{2}} \, \left[\left|f^{-2-1}\right|^{2} + \left|f_{21}\right|^{2} 
	+ \left|f^{-1-2}\right|^{2} \right. \\
	& + & \left. \left|f_{12}\right|^{2}\right] 
	\end{array}
\end{equation}
and
\begin{equation}
	\begin{array}{rcl}
	f_{2} & = & \frac{3}{(m_{e}R_A)^{2}} \, \left[\left|f^{-2}{}_{1}\right|^{2}+\left|f^{-1}{}_{2}\right|^{2}
	+ \left|f_{1}{}^{-2}\right|^{2} \right. \\
	& + & \left. \left|f_{2}{}^{-1}\right|^{2}\right]  
	\end{array}
\end{equation}
\end{subequations}
 are combinations of electron wave functions as defined in  Ref. \cite[Appendix 1]{Tomoda:1990rs}.

In the calculation of $G_{1,2}$, we have used electron radial wave functions obtained via a numerical solution of the Dirac equation with potential \cite{Kotila:2012zza, radial}
\begin{align}
	V(r) = 
	\begin{cases}
	- \alpha Z_{\rm F}\frac{3-(r/R_A)^2}{2R_A} \times \varphi(r) \mbox{ }, &r < R \mbox{ }, \\
	- \frac{\alpha Z_{\rm F}}{r} \times \varphi(r) \mbox{ },               &r \geq R \mbox{ },
	\end{cases}
\end{align}
which includes finite size corrections to the Coulomb potential of the final nucleus with charge $Z_{\rm F}$ and electron screening, due to the electronic cloud described in the Thomas-Fermi approximation by the function $\varphi(r)$. The thus obtained values of the phase-space integrals in Eq. (\ref{eqn:Gi}) are given in Table \ref{num-PSF}.
\begin{table}[htbp]
\centering
\begin{ruledtabular}
\begin{tabular}{cccc}
 & $G_{1}$ [$10^{-15}$ yr$^{-1}$] & $G_{2}$ [$10^{-15}$ yr$^{-1}$] & $Q$-value [keV]  \\
\hline
$^{76}$Ge &1.669 &1.157 & 1479.9 \\ 
$^{82}$Se &12.357 & 9.159 & 2221.4 \\
$^{130}$Te &18.464 &14.462 & 1991.4 \\
$^{136}$Xe &8.611 &6.269 & 1639.3 \\
\end{tabular}
\end{ruledtabular}
\caption{Phase-space factors $G_1$ and $G_2$ for 0$^{+}$ $\rightarrow$ 2$^{+}$ $ 0\nu\beta\beta$ decay. The $Q$-values for the above decays are reported in the last column. They are obtained by subtracting the 0$_{1}^{+}$-2$_{1}^{+}$ energy splitting in the levels of the daughter nuclei from the corresponding $Q$-values of the standard $0^{+} \rightarrow 0^{+}$ processes from Refs. \cite{Rahaman:2007ng,Redshaw:2009zz,Lincoln:2012fq,CUORE:2017tlq,CUPID:2018npf,EXO:2017poz}. }
\label{num-PSF}
\end{table}

\begin{table*}[htbp]
\centering
\begin{ruledtabular}
\begin{tabular}{lcccc}
Decay &Collaboration & $\tau_{1/2}^{0\nu, \rm exp}$ [y] &  $|\langle \lambda \rangle|$ & $|\langle \eta \rangle|$\\
\hline
$^{76}$Ge &Majorana \cite{PhysRevC.103.015501} & $>2.1\times 10^{24}$ &< $3.0\cdot10^{-4}$ & $<9.7\cdot10^{-5}$ \\
$^{82}$Se &CUPID-0 \cite{CUPID:arxiv}  & $>3.0\times 10^{23}$ & $<2.0\cdot10^{-4}$ & $<1.9\cdot10^{-4}$ \\
$^{130}$Te &Gran Sasso \cite{Arnaboldi:2002te}  & $>1.4\times 10^{23}$ & $<2.2\cdot10^{-4}$ & $<1.1\cdot10^{-4}$ \\
$^{136}$Xe &KamLAND-Zen \cite{KamLAND-Zen:2015tnh}  & $>2.6\times 10^{25}$ & $<1.7\cdot10^{-5}$ & $<1.4\cdot10^{-5}$ \\
\end{tabular}
\end{ruledtabular}
\caption{Calculated limits on the $\langle \eta \rangle$ and $\langle \lambda \rangle$ L-R model couplings. These limits are estimated by comparing our IBM-2 results for the 0$^{+}$ $\rightarrow$ 2$_1^{+}$ half-life, computed by means of Eq. (\ref{eqn:tau-lam-eta}), with the experimental limits on the $^{76}$Ge, $^{82}$Se, $^{130}$Te and $^{136}$Xe half-life (90\% C.L.).}
\label{compa}
\end{table*} 
Our results for PSFs  are comparable to those of  Ref. \cite{Tomoda:1990rs}, where the leptonic phase-space integrals were computed by making use of electron radial wave functions approximated by their leading terms in a power series expansion in $r$. As an example, 
 converted to our notation, the resulting values for $^{76}$Ge read  $G_1^{\rm pse} = 1.865 \times10^{-15}$ yr$^{-1}$ and $G_2^{\rm pse}  = 1.296 \times10^{-15}$ yr$^{-1} $. It is noteworthy that the values obtained with approximate wavefunctions are slightly larger than the values of $G_{1,2}$ reported in Table \ref{num-PSF}, as was also shown for the decays to $0^+$ states in Ref. \cite{Kotila:2012zza}.

\subsection{Limits on the $\langle \lambda 
$ and $\langle \eta \rangle$ couplings}
By combining the calculated values of the leptonic PSFs in Table \ref{num-PSF} with our IBM-2 results for the NMEs in Table \ref{num-NMEs}, we can use Eq. (\ref{eqn:tau-lam-eta}) to place limits on the $\langle \lambda \rangle$ and $\langle \eta \rangle$ couplings in L-R models. 

The upper limit on the value of the $\langle \lambda \rangle$ coupling in L-R models is obtained by setting $\langle \eta \rangle$ to zero and equating Eq. (\ref{eqn:tau-lam-eta}) to the experimental limit on the 0$^{+}$ $\rightarrow$ 2$_1^{+}$ $0\nu\beta\beta$ half-life of the mother nucleus. Analogously, by setting $\langle \lambda \rangle = 0$, one can implement the same procedure and obtain the limit on the value of the $\langle \eta \rangle$ coupling. 
The experimental results for the 0$^{+}$ $\rightarrow$ 2$_1^{+}$ $0\nu\beta\beta $ half-lives of $^{76}\mbox{Ge}$,  $^{82}\mbox{Se}$, $^{130}\mbox{Te}$ and $^{136}\mbox{Xe}$ are extracted from Refs. \cite{PhysRevC.103.015501,CUPID:arxiv,Arnaboldi:2002te,KamLAND-Zen:2015tnh}. Our upper limits on the absolute values of the $\langle \eta \rangle$ and $\langle \lambda \rangle$ L-R model parameters are reported in Table \ref{compa}.  The most stringent limits on these parameters can be set by making use of the KamLAND-Zen experimental limits on the 0$^{+}$ $\rightarrow$ 2$_1^{+}$ $0\nu\beta\beta $ half-life of $^{136}$Xe \cite{KamLAND-Zen:2015tnh},  leading to limits of  $|\langle \lambda \rangle| < 1.7\cdot10^{-5}$ and $|\langle \eta \rangle| < 1.4\cdot10^{-5}$.

\begin{figure}[htbp] 
\centering 
\includegraphics[width=8cm]{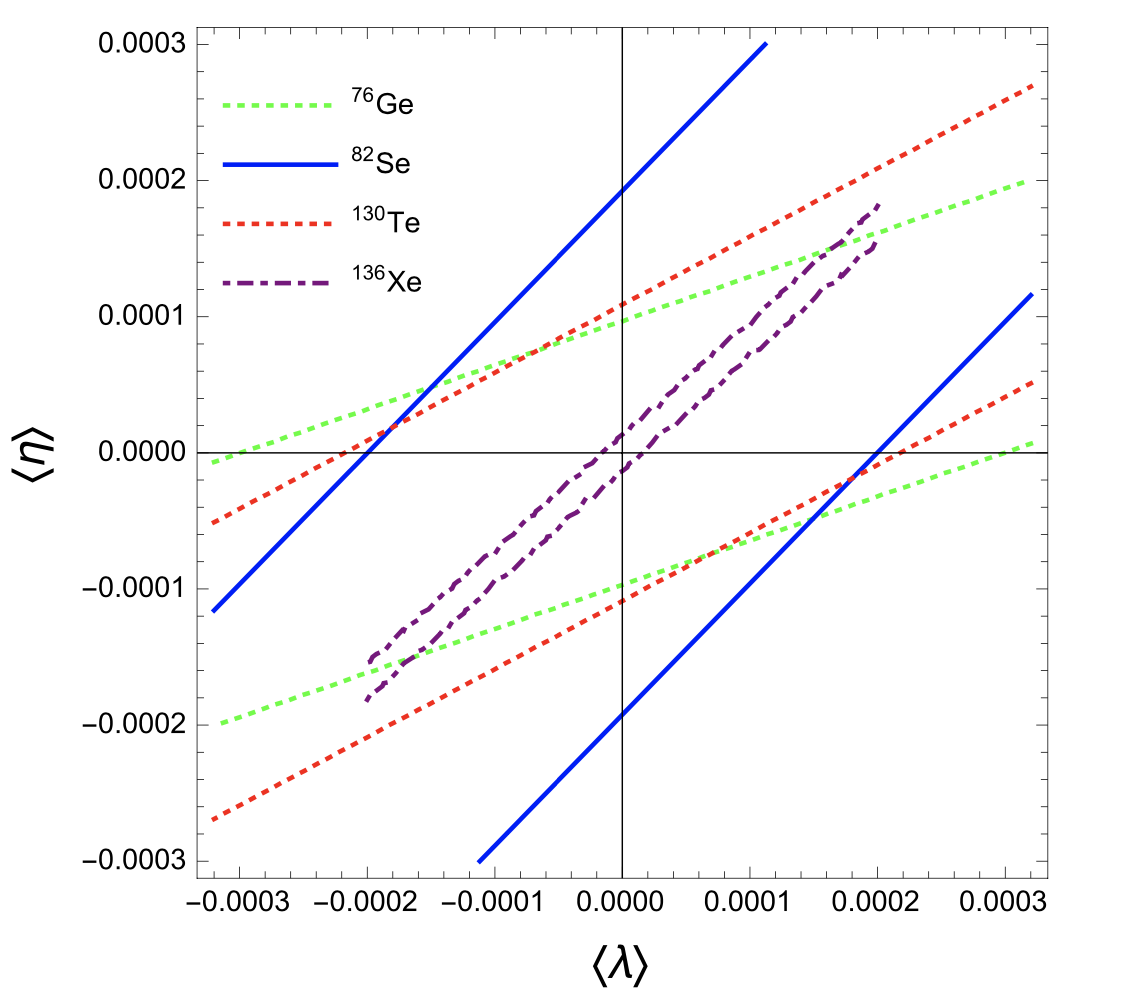}
\caption{Limits on the combination of the $\langle \lambda \rangle$ and $\langle \eta \rangle$ couplings in L-R models in the case of $^{76}$Ge (green dotted line), $^{82}$Se (continuous blue line), $^{130}$Te (red dotted line), and $^{136}$Xe (purple dot-dashed line). The parameter space outside of the parallel lines, the shaded area, is excluded.}
\label{fig1}
\end{figure}

The dependence of the parameters $\langle \eta \rangle$ and $\langle \lambda \rangle$ is also shown in Fig. \ref{fig1}, where limits on the combination of the $\langle \lambda \rangle$ and $\langle \eta \rangle$ couplings in L-R models are shown.

It is also very interesting to compare our results with those of the IBM-2 study of 0$^{+}$ $\rightarrow$ 0$^{+}$ $0\nu\beta\beta$ decays.
While one can set even more stringent limits on $\langle \eta \rangle$ and $\langle \lambda \rangle$ in the case of 0$^{+}$ $\rightarrow$ 0$^{+}$ $0\nu\beta\beta$ transitions \cite{Kotila:2021xgw}, namely $\sim 10^{-9}$ for $\langle \eta \rangle$ and $\sim10^{-7}$ for $\langle \lambda \rangle$, in this case one cannot disentangle the $\langle \eta \rangle$ and $\langle \lambda \rangle$ dependencies from their dependence on the $\langle m_\nu \rangle$ parameter. This is one of the reasons why 0$^{+}$ $\rightarrow$ 2$_1^{+}$ $0\nu\beta\beta$ decay searches are worth investigating and why 0$^{+}$ $\rightarrow$ 2$_1^{+}$ experimental searches are conducted in parallel with those for standard 0$^{+}$ $\rightarrow$ 0$^{+}$ transitions.
 
\section{Conclusions}
\label{sec:Conc} 
We have computed the neutrinoless double-$\beta$ ($0\nu\beta\beta$) decay nuclear matrix elements between the ground state and the first excited $2^+$ state of $^{76}\mbox{Ge} \rightarrow {}^{76}\mbox{Se}$,  $^{82}\mbox{Se} \rightarrow{}^{82}\mbox{Kr}$, $^{130}\mbox{Te} \rightarrow {}^{130}\mbox{Xe}$ and $^{136}\mbox{Xe} \rightarrow {}^{136}\mbox{Ba}$ within the framework of the microscopic Interacting Boson Model (IBM-2) \cite{Arima:1977vie,Otsuka:1978zz,Barea:2009zza,Iachello:2006fqa} by considering the exchange of a Majorana neutrino between two nucleons ($2N$-mechanism) \cite{Doi:1985dx,Tomoda:1999zc}.
The IBM-2 formalism was widely used in the past to obtain results for nuclear observables, including the spectrum, the electromagnetic and the weak decays, but also to explore the possible emergence of beyond-the-Standard Model effects in the weak interactions of nuclei; see e.g., Refs. \cite{Barea:2009zza,Barea:2015kwa,Santopinto:2018nyt,Graf:2018ozy,Deppisch:2020ztt}. Our results for the $^{76}$Ge(0$_{1}^{+}$) $\rightarrow$ $^{76}$Se(2$_{1}^{+}$) NMEs stand in between the PHFB results and QRPA results from the literature. For other reported cases, to our knowledge, this is the first calculation.

We have also calculated the relevant leptonic phase-space integrals numerically by making use of exact Dirac wave functions with finite nuclear size and electron screening \cite{Kotila:2012zza} in order to set some limits on the standard couplings of L-R models, $|\langle \lambda \rangle|$ and $|\langle \eta \rangle|$. As in the case of decays to $0^+$ states, the PSFs are found to be slightly smaller than previous values obtained with approximate wavefunctions. 

The most stringent limits on the parameters $|\langle \lambda \rangle|$ and $|\langle \eta \rangle|$ can be obtained from the 0$^{+}$ $\rightarrow$ 2$_1^{+}$ $0\nu\beta\beta $ half-life of   $^{136}$Xe \cite{KamLAND-Zen:2015tnh}, leading to $|\langle \lambda \rangle| < 1.7\cdot10^{-5}$ and $|\langle \eta \rangle| < 1.4\cdot10^{-5}$.
While one can set even more stringent limits on $\langle \eta \rangle$ and $\langle \lambda \rangle$ in the case of 0$^{+}$ $\rightarrow$ 0$^{+}$ $0\nu\beta\beta$ transitions, in this case one cannot disentangle the $\langle \eta \rangle$ and $\langle \lambda \rangle$ dependencies from their dependence on the $\langle m_\nu \rangle$ parameter, making  0$^{+}$ $\rightarrow$ 2$_1^{+}$ $0\nu\beta\beta$ decay searches worth investigating further.

\begin{acknowledgments}
This work was supported by the Academy of Finland, Grant No. 314733, 320062, 345869, and INFN, Italy.
\end{acknowledgments}


\end{document}